\documentclass[english]{cccconf}
\usepackage[comma,numbers,square,sort&compress]{natbib}
\usepackage{epstopdf}
\usepackage{booktabs} % 必须的宏包，用于横线
\usepackage{multirow} % 用于合并行
\usepackage{stfloats}
\usepackage{graphicx}
\usepackage{subcaption}
\usepackage{amsmath}
\usepackage{amssymb}
\usepackage{siunitx}
\begin{document}

\title{DCAF-Net: Dual-Channel Attentive Fusion Network for Lower Limb Motion Intention Prediction in Stroke Rehabilitation Exoskeletons}

\author{Liangshou Zhang\aref{amss},
        Yanbin Liu\aref{hit},
        Hanchi Liu\aref{amss},
        Zheng Sun\aref{amss},
        Haozhi Zhang\aref{inspur}*,
        Yang Zhang\aref{hit}*,
        Xin Ma\aref{amss}*}

\affiliation[amss]{Center for Robotics, School of Control Science and Engineering, Shandong University, Jinan, 250061, China\email{maxin@sdu.edu.cn}}
\affiliation[hit]{Rehabilitation and Physical Therapy Department, Affiliated Hospital of Shandong University of Traditional Chinese Medicine, Jinan, 250011, China\email{zhangyang982003@163.com}}
\affiliation[inspur]{INSPUR GROUP CO.,LTD, Jinan, 250101, China.}

\maketitle

\begin{abstract}
Rehabilitation exoskeletons have shown promising results in promoting recovery for stroke patients. Accurately and timely identifying the motion intentions of patients is a critical challenge in enhancing active participation during lower limb exoskeleton-assisted rehabilitation training. This paper proposes a Dual-Channel Attentive Fusion Network (DCAF-Net) that synergistically integrates pre-movement surface electromyography (sEMG) and inertial measurement unit (IMU) data for lower limb intention prediction in stroke patients. First, a dual-channel adaptive channel attention module is designed to extract discriminative features from 48 time-domain and frequency-domain features derived from bilateral gastrocnemius sEMG signals. Second, an IMU encoder combining convolutional neural network (CNN) and attention-based long short-term memory (attention-LSTM) layers is designed to decode temporal-spatial movement patterns. Third, the sEMG and IMU features are fused through concatenation to enable accurate recognition of motion intention. Extensive experiment on 11 participants (8 stroke subjects and 3 healthy subjects) demonstrate the effectiveness of DCAF-Net. It achieved a prediction accuracies of 97.19\% for patients and 93.56\% for healthy subjects. This study provides a viable solution for implementing intention-driven human-in-the-loop assistance control in clinical rehabilitation robotics.
\end{abstract}


% Please remove or comment out the following line if the footnote is not necessary
\footnotetext{This work was supported in part by National Key Research and Devel opment Program 2023YFB4706104.}

\section{Introduction}

According to epidemiological survey data released by the Chinese Stroke Association, stroke has emerged as the leading cause of mortality among Chinese residents. Current statistics indicate that approximately 14.94 million stroke survivors live in China, with 3.3 million new cases and 1.54 million stroke-related deaths reported annually. Notably, about 80\% of survivors experience varying degrees of long-term disability. Among the sequelae of stroke, motor issues such as limb weakness, muscle rigidity, and movement incoordination are particularly prevalent\cite{jiangyj,xurui}.

While lower limb exoskeletons have demonstrated efficacy in facilitating gait rehabilitation for stroke patients, most existing systems rely on passive control paradigms that enforce predetermined movement trajectories\cite{Meng}. This rigid approach often leads to user fatigue and psychological resistance during rehabilitation training. Consequently, developing intention-driven adaptive control strategies that can accurately decode user motion intent and dynamically adjuste exoskeleton assistance is a crucial research frontier. 

sEMG and IMU has emerged as prominent research focuses for intention recognition in the field of exoskeleton human-robot interaction\cite{wangdj}. sEMG detects neuromuscular activation 30-150 ms before movement onset, making it widely used for motion intention recognition\cite{Zhngyuepeng,weizj}. Anselmino et al. used a controller based on two Support Vector Machines exploiting four EMG signals of the thigh muscles to detect the Toe-off and the direction of the upcoming step, achieving an  accuracy of 92.91\%\cite{eugenio}. Niu et al. employed an Improved Dense Convolutional Network model to recognize motion intentions in the affected hand of amputees based on their sEMG signals, achieving an accuracy of 94.50\%\cite{sunj}. IMU has gained traction due to its robustness and cost-effectiveness in capturing kinematic signals\cite{chenyilian}. Zhang et al. proposed a Motion Forecasting Network (MoFCNet) for 3D pose prediction based on historical IMU data, achieving a mean joint rotation error of 10.34°\cite{zhangx}. Xuan et al. used a dynamic time warping (DTW) algorithms on a dedicated IMU dataset to recognize the real-time lower-limb motion\cite{XuanL}. However, these methods rely solely on a single modality, which may lead to poor robustness in recognition.
% Nevertheless, IMU-derived physical signals inherently lag behind neuromuscular activity, because kinetics data has the hysteretic nature\cite{wangxj}.

In recent years, the fusion of IMU and EMG has garnered increasing attention. Li et al. used a multistream convolutional neural network(CNN) architecture with a fine-tuning transfer strategy based on the sEMG and IMU signals to recognize the gestures, achieving an accuracy of 98.12\% and a latency of 103 ms for real-time recognition\cite{ligy}.  Han et al. employed a CNN with bidirectional long short-term memory based on the sEMG and IMU to predict the elbow joint angle, achieving an RMSE of 2.0345\cite{hanjd}. However, theses studies have mainly focused on motion recognition and joint angle prediction either during or after the completion of movements. However, there has been no research exploring the use of pre-motion signals to predict movement intentions for active control of exoskeletons.

 Aiming to address the above challenge, this study proposes the DCAF-Net, which integrates pre-movement EMG-IMU multimodal data from stroke patients to predict lower-limb motion intent. DCAF-Net was evaluated on a cohort of 8 stroke patients and 3 healthy participants. The model achieved an average prediction accuracy of 97.19\% for stroke patients and 93.56\% for healthy individuals. These results suggest that DCAF-Net can effectively decode pre-motion intent in neurologically impaired patients. Importantly, the high accuracy and low latency (sub-200 ms) of DCAF-Net lay the foundation for active exoskeleton control, which is expected to enhance active patient engagement during rehabilitation training.

\section{Data And Model Structure}

\subsection{Paradigm}

\begin{figure}[h]
  \centering
  \includegraphics[width=0.45\textwidth]{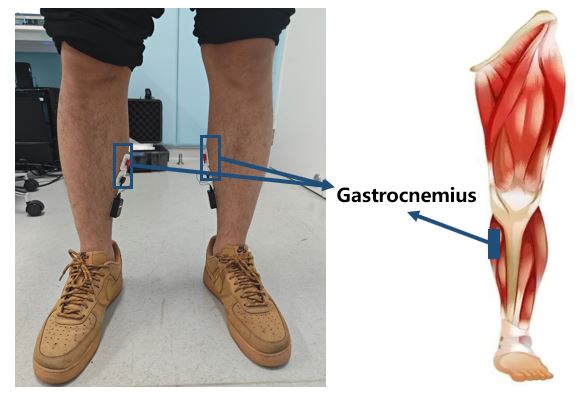}
  \caption{The position of sEMG electrodes}
  \label{fig1}
\end{figure}

In the prediction of lower-limb stepping motion intention, sEMG signals and IMU data were collected from participants' lower limbs. The sEMG signals were acquired using the NeuroHUB wireless surface electromyography module, a multi-channel, wireless telemetry EMG acquisition system independently developed by Neuracle Corporation (China). Each sensor integrates one sEMG channel, a triaxial accelerometer, a triaxial gyroscope, and a triaxial magnetometer. The device operates at a sampling rate of 2,000 Hz for sEMG signals and 200 Hz for IMU data. Considering the critical role of the gastrocnemius muscle in maintaining postural balance and propelling movement\cite{tz}, sEMG signals and IMU data were specifically collected from the gastrocnemius muscles of both legs. The sensors placement is illustrated in Figure \ref{fig1}.

The experiment involved 11 participants, including 8 stroke subjects(Sub1-Sub8, 32-77 years old) and 3 healthy subjects(Sub9-Sub11, 20-27 years old). Each subject performed four stepping tasks following verbal instructions: small-step (20 cm) with the right leg, small-step (20 cm) with the left leg, large-step (40 cm) with the right leg, and large-step (40 cm) with the left leg. Each task was repeated 50 times. sEMG and IMU data were recorded from movement initiation to completion.

\subsection{Signal Preprocessing}
Signal preprocessing includes signal filtering, signal extraction, data segmentation.

The raw sEMG signals conclude muscle electrical activity as well as unavoidable noise, such as electronic noise from the amplification system. Therefore, filtering is required to remove baseline noise before further analysis. A Butterworth bandpass filter with a frequency range of 10 Hz to 500 Hz was applied to reduce the impact of high-frequency noise, and a 50 Hz Butterworth notch filter was used to mitigate powerline interference.

For the IMU data, the vertical ground reaction acceleration was utilized. Since the IMU data is susceptible to high-frequency noise, a 4th-order Butterworth low-pass filter was applied to the acceleration signal to remove high-frequency noise.

\begin{figure}[htb]
  \centering
      \begin{subfigure}[b]{0.4\textwidth}
    \includegraphics[width=\textwidth]{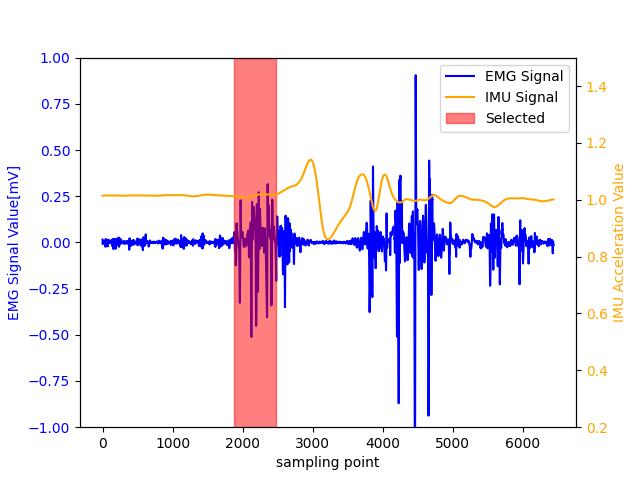}
    \caption{Right leg signal}
    \label{fig:r20r}
  \end{subfigure}
        \begin{subfigure}[b]{0.4\textwidth}
    \includegraphics[width=\textwidth]{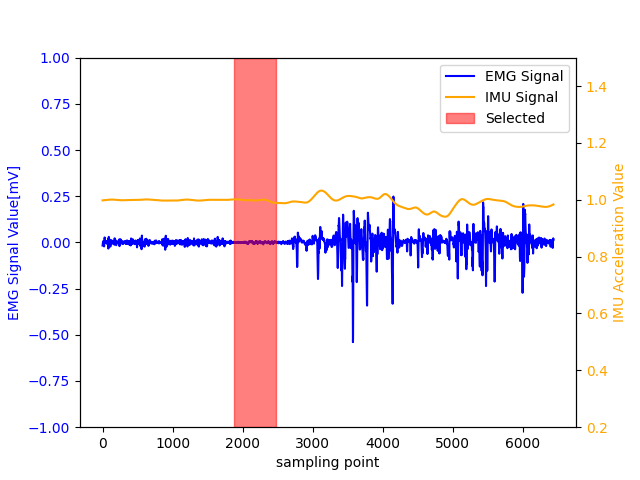}
    \caption{Left leg signal}
    \label{fig:r20l}
  \end{subfigure}
  \caption{Extraction of pre-movement sEMG}
  \label{fig2}
\end{figure}

Since the goal is to predict the participants' motion intention for active exoskeleton control and the patients were immobilized in the exoskeleton, they were unable to complete the stepping motion. Therefore, in the experiment, sEMG and IMU data from the 300 milliseconds preceding heel-off were extracted. This 300 ms window of data was used to predict the stepping leg (left or right) and step size (small or large). Heel-off detection was performed using the method described in \cite{mikko}, where the moment when the vertical acceleration reached 1 m/s² was identified as the heel-off event. A schematic diagram of the data extraction process is shown in Figure \ref{fig2}. The blue segments represent sEMG signals, yellow segments denote sEMG signals, and red segments indicate the pre-movement sEMG signals selected for analysis.

The raw 300 ms sEMG/IMU segments were divided into overlapping sub-windows for model training. A window length of 200 ms was adopted, as it aligns with the temporal scale required for motion intent decoding\cite{eugenio}. Using a 200 ms window length and a 50 ms sliding step, each segment generates three sub-windows. Each sub-window inherits the label of its parent segment, enhancing the generalization capability of the recognition model.

\subsection{Feature Extraction}

To accurately predict motion intention, a comprehensive set of 48 discriminative features was extracted from the EMG signals, encompassing three domains: time-domain, frequency-domain, and time-frequency representations. These features were designed to capture neuromuscular activation patterns, energy distribution, and transient dynamics. A complete feature list is provided in Table \ref{tab:feature_summary}.The detailed descriptions of each feature are provided in \cite{sidi}.

Therefore, for each 200 ms segment of the EMG signal, comprising a total of 400 sampled data points, 48 features are extracted. Simultaneously, the 200 ms segment of the IMU vertical acceleration data, consisting of 40 sampled data points, forms the input to our network.

\begin{table}[htbp]
\centering
\caption{EMG Signal Feature}
\label{tab:feature_summary}
\begin{tabular}{cl}
\hhline
\textbf{Domain Category} & \textbf{Feature Name} \\
\hline
\multirow{20}{*}{Time Domain}
& MAV (Mean Absolute Value) \\
& WL (Waveform Length) \\
& Modified Waveform Length \\
& SSC (Slope Sign Changes) \\
& ZC (Zero Crossings) \\
& IEMG (Integrated EMG) \\
& SSI (Simple Square Integral) \\
& MEAN \\
& VAR (Variance) \\
& SD (Standard Deviation) \\
& RMS (Root Mean Square) \\
& WAMP (Willison Amplitude) \\
& Myoelectric Percentage \\
& Skewness \\
& Kurtosis \\
& Difference Absolute Std \\
& MMAV1/MMAV2  \\
& MAVSLP (MAV Slope) \\
& V-Order \\
& Average Amplitude Change \\
& Max \\
& Min \\
& Crest Factor \\
& Mean Absolute Deviation \\
& Energy Entropy \\ \hline
\multirow{10}{*}{Frequency Domain}

& MNF (Mean Frequency) \\
& MDF (Median Frequency) \\
& MNP (Mean Power) \\
& TTP (Total Power) \\
& PKF (Peak Frequency) \\
& PSR (Power Spectrum Ratio) \\
& AR Coefficients \\
& SMR (Spectral Moment Ratio) \\
& FCV (Frequency Center Variance) \\
& Mean Power Spectral Density \\ \hhline

\end{tabular}
\end{table}

\subsection{Dual-Channel Attentive Fusion Network}

\begin{figure*}[t]
  \centering
  \includegraphics[width=0.95\textwidth]{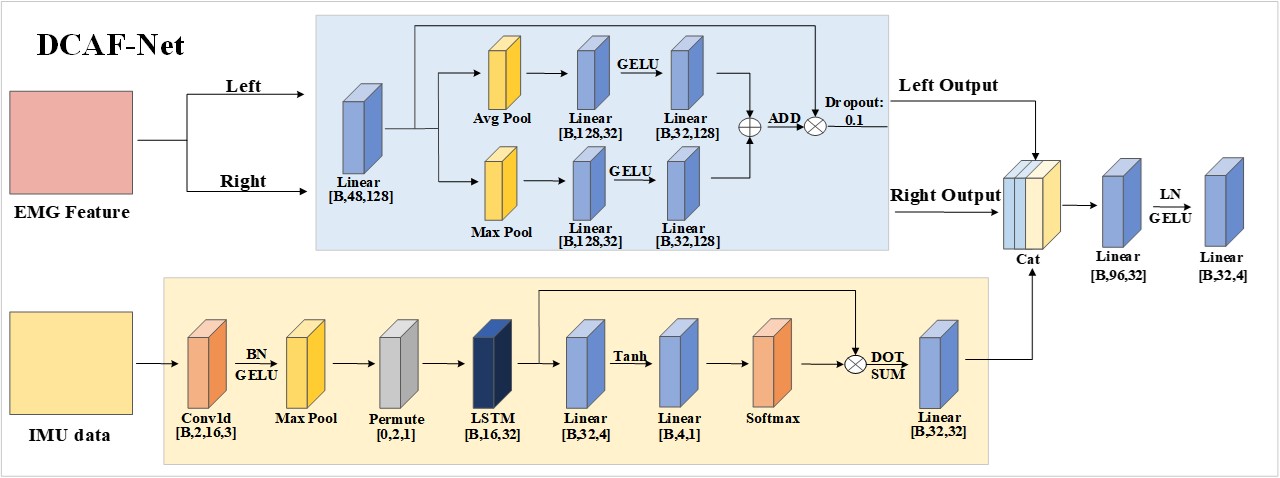}
  \caption{The structure of DCAF-Net}
  \label{fig3}
\end{figure*}

The Dual-Channel Attentive Fusion Network (DCAF-Net) is a multimodal deep learning framework designed for robust gait intention prediction by sEMG and IMU signals. The network architecture is shown in Figure \ref{fig3}. The architecture addresses the challenges of temporal variability in motion patterns and inter-channel signal heterogeneity through three specialized modules.

1) \textbf{sEMG Feature Extraction with Adaptive Attention}

Each sEMG channel is processed by a dedicated subnetwork that begins with a linear projection of 48 handcrafted features into a 128-dimensional latent space. To mitigate the impact of non-discriminative muscle activation patterns, a channel attention mechanism dynamically reweights feature importance. This module combines global average and max pooling to generate channel-wise attention weights, enhancing sensitivity to biomechanically relevant biomarkers. The refined features are then compressed into a 64-dimensional representation through dropout-regularized linear layers. The channel attention mechanism is formulated as follows:
\begin{equation}
\mathbf{f}_{\text{avg}} = \text{AvgPool}(\mathbf{f}) \in \mathbb{R}^C
\end{equation}
\begin{equation}
\mathbf{f}_{\text{max}} = \text{MaxPool}(\mathbf{f}) \in \mathbb{R}^C
\end{equation}
\begin{equation}
\mathbf{w} = \sigma\left(\mathcal{F}_{\text{MLP}}(\mathbf{f}_{\text{avg}}) + \mathcal{F}_{\text{MLP}}(\mathbf{f}_{\text{max}})\right) \in \mathbb{R}^C
\end{equation}
\begin{equation}
\mathbf{f}_{\text{attn}} = \mathbf{f} \odot \mathbf{w}
\end{equation}
where \( \mathcal{F}_{\text{MLP}} \) denotes a two-layer MLP with GELU activation, and \( \sigma \) is the sigmoid function. This mechanism ensures that the network focuses on the most discriminative EMG features while suppressing noise and artifacts.

2) \textbf{Attention-Driven Hierarchical IMU Temporal Encoding}

The IMU branch processes dual-channel sequential data (40 time steps per channel) through a hybrid temporal modeling pipeline. A 1D convolutional layer first extracts local kinematic patterns , followed by max pooling for temporal downsampling. LSTM layers then capture long-range dependencies across gait cycles. To address the temporal challenges in IMU data, the attention-based pooling mechanism dynamically computes adaptive weights for each time step, emphasizing critical motion phases such as the heel-strike transition. The final IMU representation is obtained by aggregating time-aware features through weighted summation. The attention-based pooling is defined as:
\begin{equation}
\alpha_t = \text{Softmax}\left(\mathbf{v}^\top \tanh(\mathbf{W}_h \mathbf{h}_t + \mathbf{b})\right)
\end{equation}
\begin{equation}
\mathbf{c} = \sum_{t=1}^T \alpha_t \mathbf{h}_t
\end{equation}
where \( \mathbf{h}_t \in \mathbb{R}^{32} \) is the LSTM hidden state at time \( t \), and \( \mathbf{W}_h \), \( \mathbf{v} \), \( \mathbf{b} \) are learnable parameters. This approach ensures that the network dynamically focuses on the most informative time steps, improving robustness to temporal variations.

3) \textbf{Multimodal Fusion and Classification}

The network combines sEMG features (64 dimensions per limb) and IMU embeddings (32 dimensions) into a unified 160-dimensional representation via concatenation. A lightweight classifier, composed of layer normalization and GELU-activated linear layers, transforms the fused features into motion intention probabilities. This architecture effectively preserves modality-specific information while promoting implicit cross-modal interactions in deeper layers. The classification process is formally defined as:
\begin{equation}
P(y|\mathbf{x}) = \text{Softmax}\left(\mathbf{W}_2 \cdot \text{GELU}(\mathbf{W}_1[\mathbf{f}_L; \mathbf{f}_R; \mathbf{f}_{\text{IMU}}])\right)
\end{equation}

where \( \mathbf{W}_1 \in \mathbb{R}^{128 \times 160} \), \( \mathbf{W}_2 \in \mathbb{R}^{4 \times 128} \). This design preserves modality-specific information while enabling implicit cross-modal interaction in deeper layers.
%area. Captions should appear below graphical objects, as in Fig.~\ref{fig1}.

\section{Result And Discussion}
The DCAF-Net framework was implemented on an Ubuntu 20.04 system with NVIDIA GeForce RTX 4090 GPU acceleration, utilizing PyTorch 2.30 for model development. A cohort of 11 participants (8 stroke subjects and 3 healthy subjects) was enrolled in this experiment. 

The experimental protocol involved 200 gait cycles per subject during data acquisition. Through sEMG signal window segmentation, 600 signal windows were extracted from each participant. The dataset was partitioned into training and test sets at an 8:2 ratio, resulting in 480 training samples and 120 test samples per participant.

As detailed in Table \ref{tab:results}, the classification results demonstrate the performance of DCAF-Net across all participants, including eight stroke patients (Subjects 1-8) and three healthy individuals (Subjects 9-11). Among the stroke patients, the group achieved an average accuracy of 96.2\%, with four individuals (Subjects 1, 2, 5, and 6) attaining perfect accuracy of 100\%. The overall mean accuracy for the stroke patient group was 97.19\%. In comparison, the healthy control group achieved a slightly lower average accuracy of 93.56\%. These results highlight the effectiveness of DCAF-Net in motion intention prediction across both stroke patients and healthy individuals.

\begin{table}[htbp]
\centering
\caption{Classification performance of DCAF-Net}
\label{tab:results}
\begin{tabular}{c l S[table-format=3.2]}
\hhline
\textbf{Subject} & \textbf{Affected Limb} & \textbf{Accuracy (\%)} \\ 
\hline
1  & Left   & 100.00 \\
2  & Right  & 100.00 \\
3  & Right  & 96.67 \\
4  & Left   & 99.17 \\
5  & Right  & 100.00 \\
6  & Right  & 100.00 \\
7  & Left   & 91.67 \\
8  & Right  & 90.00 \\
\hline
9  & Healthy & 90.67 \\
10 & Healthy & 93.33 \\
11 & Healthy & 96.67 \\
\hhline
\end{tabular}
\end{table}

\begin{figure}[hbt]
  \centering
      \begin{subfigure}[b]{0.5\textwidth}
    \includegraphics[width=\textwidth]{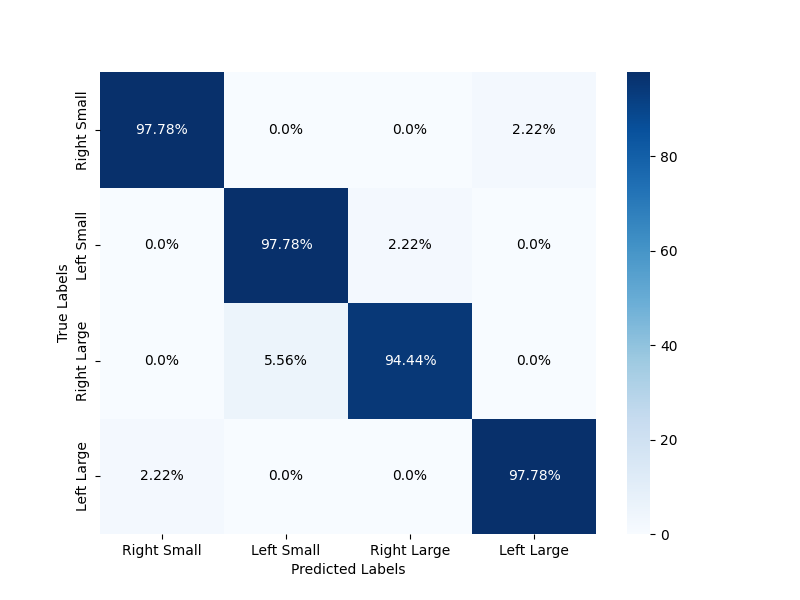}
    \caption{Left Leg}
    \label{fig:IMU}
  \end{subfigure}

  \begin{subfigure}[b]{0.5\textwidth}
  \includegraphics[width=\textwidth]{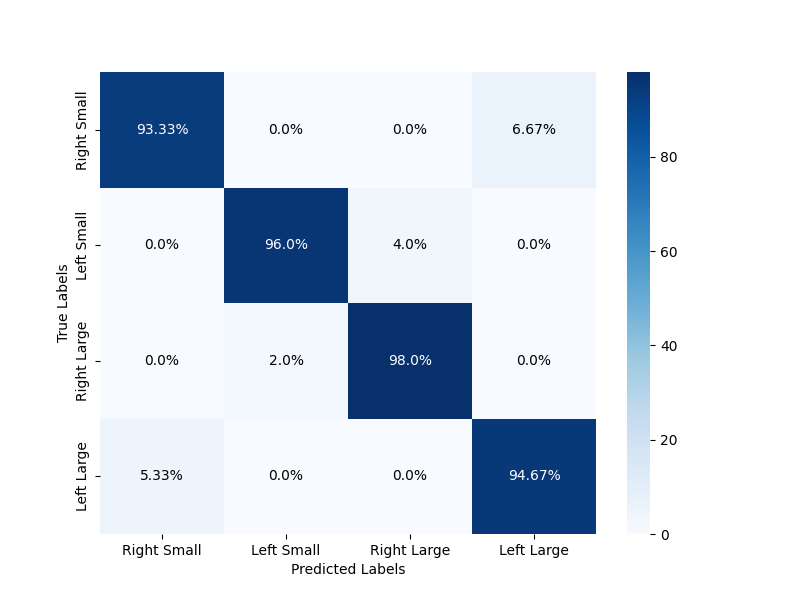}
  \caption{Right Leg}
  \label{fig:EMG}
  \end{subfigure}
  \caption{Confusion matrixes of the prediction for patients}
  \label{mix}
\end{figure}

\begin{figure}[htb]
  \centering
  \includegraphics[width=0.5\textwidth]{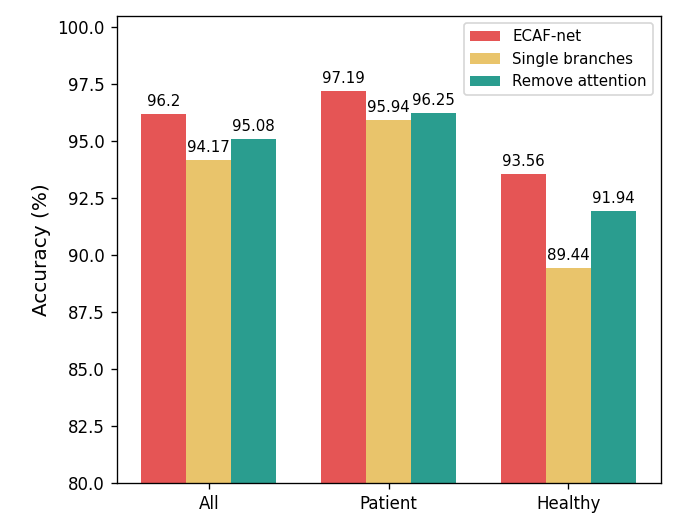}
  \caption{Ablation experiment}
  \label{Ablation}
\end{figure}

To investigate the impact of different affected limbs on model predictions, Figure \ref{mix} presents the confusion matrices for stroke patients with different affected limbs (left leg and right leg). As shown in the figure, the confusion matrices for left-leg patients and right-leg patients exhibit similar patterns, with comparable proportions and types of misclassifications. This suggests that the difference in the affected limb does not significantly influence the model's predictive performance, indicating that the model demonstrates consistent robustness across patients with different affected limbs.

Figure \ref{Ablation} presents the ablation experiments of DCAF-Net. As shown in the figure, when the model transitions from a dual-branch structure to a single-branch structure, its ability to extract features from electromyographic (sEMG) signals diminishes. Specifically, the average accuracy of motion intention prediction decreases by 1.65\% for healthy individuals and 4.12\% for stroke patients. Furthermore, upon removing the attention mechanisms from both the EMG feature extraction module and the IMU feature extraction module, the average accuracy of motion intention prediction declines by 0.94\% for healthy individuals and 1.62\% for stroke patients. This demonstrates the effectiveness of the dual-branch module and attention mechanisms in DCAF-Net.

\begin{figure}[t]
  \centering
  \includegraphics[width=0.5\textwidth]{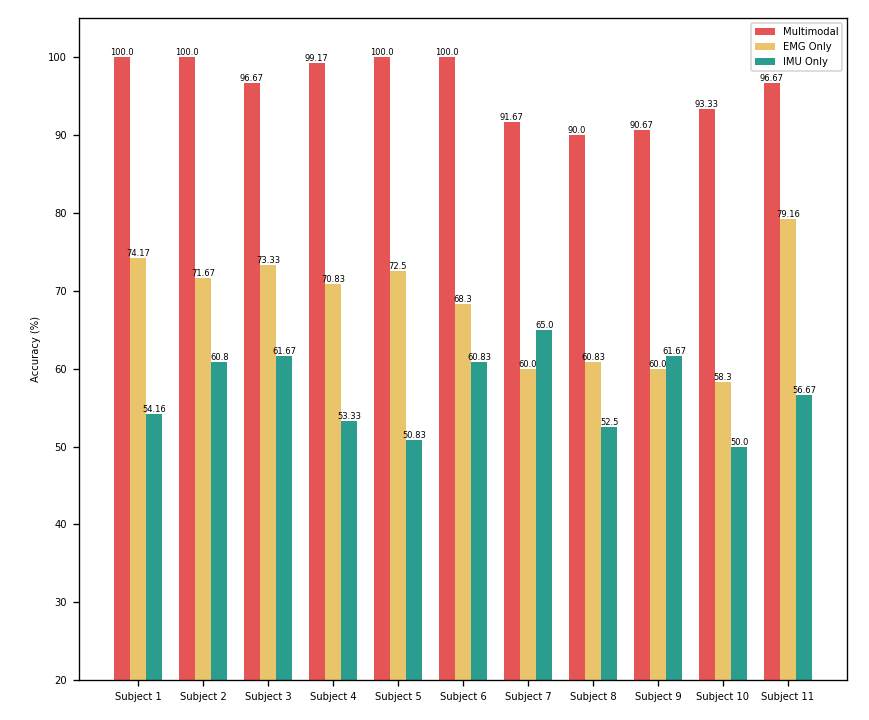}
  \caption{Comparison of performance metrics for multimodal and single-modal approaches}
  \label{Comparison}
\end{figure}

To investigate the role of each modality in the task of motion intention prediction, Figure \ref{Comparison} illustrates the prediction accuracy of single-modality (IMU-only and EMG-only) and dual-modality fusion across eleven subjects, including eight stroke patients and three healthy individuals. The results demonstrate that the fused modality achieves superior accuracy compared to any single modality for all subjects. Specifically, the IMU-only approach yields an average accuracy of 57.39\% for stroke patients and 56.11\% for healthy individuals in motion prediction, while the EMG-only method achieves 68.95\% for stroke patients and 65.82\% for healthy individuals. In contrast, the fusion of IMU and EMG data significantly improves the average accuracy to 96.96\% for stroke patients and 93.56\% for healthy individuals, highlighting the complementary advantages of multi-sensor integration.

\section{Conclusion}

 This paper presents DCAF-Net, a novel multimodal fusion framework that synergistically integrates neuromuscular pre-activation (sEMG) and kinematic precursors (IMU) for lower-limb motion intent prediction in stroke rehabilitation. First a dual-channel adaptive channel attention module is designed to extract discriminative features from 48 time-domain and frequency-domain features derived from bilateral gastrocnemius sEMG signals. Second, an IMU encoder combining CNN and attention-LSTM layers is designed to decode temporal-spatial movement patterns. Third, the sEMG and IMU features are fused through concatenation to enable accurate recognition of motion intention. Validated through experiments involving 3 healthy participants and 8 stroke patients, the proposed method achieved an average prediction accuracy of 97.19\% in the patient cohort and 93.56\% for healthy subjects. This demonstrates its effectiveness in lower-limb motion intent prediction and provides a viable solution for intention-driven human-in-the-loop assistance control in rehabilitation exoskeletons.

\end{document}